\documentclass{ws-ijmpd}
\usepackage[super,compress]{cite}
\begin{document}

\markboth{John Ellis}
{Time to move on?}

%
\catchline{}{}{}{}{}
%

\title{TIME TO MOVE ON?
}

\author{JOHN ELLIS\footnote{Also
Theoretical Physics Department, CERN, CH-1211 Geneva 23, Switzerland.}}

\address{Theoretical Particle Physics and Cosmology Group, Physics Department, 
King's College London, London WC2R 2LS, United Kingdom\\
John.Ellis@cern.ch}



\maketitle


\begin{abstract}
Cosmology and particle physics have long been dominated by theoretical paradigms:
Einstein's general theory of relativity in cosmology and the Standard Model of particle
physics. The time may have come for paradigm shifts. Does cosmological inflation require a modification of Einstein's gravity?
Have experiments at the LHC discovered a new particle beyond the Standard Model?
It is premature to answer these questions, but we theorists can dream about the possibilities.
\end{abstract}


\ccode{PACS numbers: 04.25.dg, 04.30.Nk, 04.65.+e, 12.0.Dm, 12.60.Jv. 95.30.Sf, 98.80.Es \\
~\\
KCL-PH-TH/2016-07, LCTS/2016-05, CERN-TH-2016-036}


\section{Introduction}	

It is a privilege to have been invited to give this talk at the end of a meeting celebrating the centenary of Einstein's
general theory of relativity. The organisers undoubtedly expected me to talk about models of cosmological
inflation, and most of this talk will indeed be about this subject, and whether it suggests a modification of the minimal
Einstein-Hilbert action such as an $R^2$ term~\cite{Staro} or supergravity~\cite{noscale}.
However, during this conference we received dramatic news from CERN of
hints of the possible existence of what might, perhaps be a new massive particle beyond the Standard Model~\cite{X}.
In the words of the CERN Director-General, Fabiola Gianotti, we theorists "are allowed to be slightly excited"~\cite{FG},
and I will try to share some of this excitement at the end of my talk~\footnote{The
discovery of gravitational waves was announced~\cite{GW} while I was writing up this talk, and I have added a note on 
constraints this discovery imposes on
possible modifications of Einstein's theory.}.

\section{Cosmological Inflation}

For almost two decades now, the cosmological paradigm has been the $\Lambda$CDM model,
based on the general theory of relativity with the addition of a cosmological constant $\Lambda$.
Einstein supposedly regarded the introduction in 1917 of $\Lambda$ as ``the biggest
blunder of his life"~\cite{blunder}, but nowadays we would beg to differ. Its presence is a possibility allowed by
the general principles of relativity, and hence is probably compulsory. The present-day
cosmological constant has depressing implications for astronomers in the far future, who
will only be able to study our local gravitationally-bound group of galaxies. However, according
to the hypothesis of cosmological inflation~\cite{inflation}, the energy density of the very early universe may
have been dominated by an (almost) constant term that would have generated an epoch of
(near-) exponential expansion that could explain the great size and age of the universe today
and, as an added bonus, have been the origin of the structures seen in the universe today,
from clusters of galaxies to Donald Trump.

The most important evidence for cosmological inflation, and constraints on models, come from
observations of fluctuations in the cosmic microwave background radiation (CMB), 
e.g., by the Planck satellite~\cite{Planck}. Together with other
cosmological and astrophysical data, these CMB measurements are consistent with the cosmological 
concordance model with the following energy densities as fractions of that of a universe that is
geometrically flat: $\Omega_\Lambda \simeq 0.69,  \Omega_{\rm matter} \simeq 0.31$, with most
of the latter being cold dark matter (CDM)~\cite{pdg}. I now discuss aspects of the inflationary paradigm and
how it is constrained by CMB data.

\subsection{Slow-Roll Inflationary Models}

Most interpretations of the CMB data adopt the framework of slow-roll inflationary models~\cite{inflation},
in which the near-exponential expansion of the universe would have been driven by the almost constant field energy
$V$ of some inflaton scalar field $\phi$:
\begin{equation}
\epsilon \; \equiv \; \frac{1}{2} M_P^2 \left( \frac{V^\prime}{V} \right)^2, \; \eta \; \equiv \; M_P^2 \left( \frac{V^{\prime \prime}}{V} \right) \; \ll \; 1 \, .
\label{epsiloneta}
\end{equation}
Quantum fluctuations in $\phi$ would have led to scalar and tensor perturbations in the CMB,
and the principal observables are the power-law tilt in the scalar spectrum $n_s$, and the tensor-to-scalar
ratio $r$, which are given in the slow-roll approximation by
\begin{equation}
n_s \; = \; 1 - 6 \epsilon + 2 \eta, \; r \; =\; 16 \epsilon \, ,
\label{observables}
\end{equation}
respectively. It is a generic prediction of slow-roll models that the CMB fluctuations should
be Gaussian to a very good approximation.

\begin{figure}[h!]
\vspace{-6mm}
\centering
	\scalebox{0.4}{\includegraphics{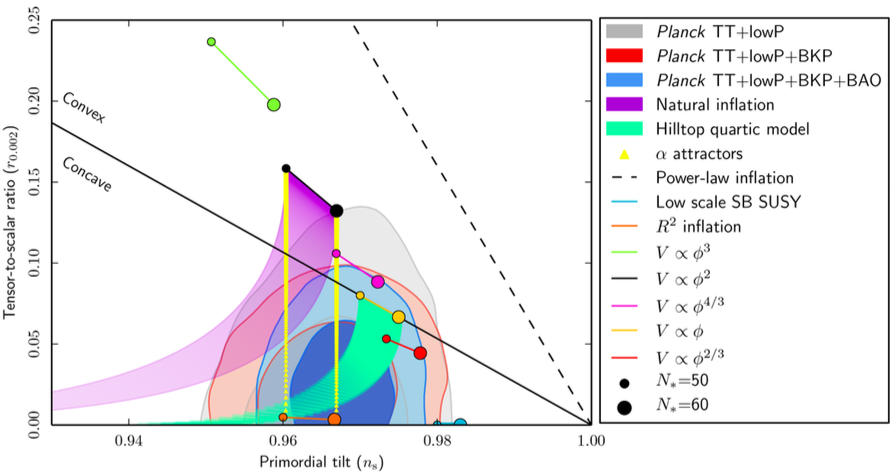}} 
		\caption{The marginalised joint 68 and 95\% CL regions for the tilt in the CMB scalar perturbation spectrum,
		$n_s$, and the relative magnitude of the tensor perturbations, $r$, obtained from the Planck 2015 data
		and their combinations with BICEP2/Keck Array and/or BAO data~\protect\cite{combined}, confronted
		with the predictions of some inflationary models.}
		\label{fig:Planck2015}
		\label{fig:nsr}
\vspace{-5mm}
\end{figure}

There was great excitement when the BICEP2 experiment reported~\cite{BICEP2} the measurement of
B-mode polarisation in the CMB, which might have been generated by primordial tensor
(quantum gravitational-wave) perturbations. However, these are now thought to consistent with
pollution by dust~\cite{Planckdust}, and a recent combined analysis of BICEP2/Keck array and Planck data
yields only an upper limit on $r < 0.1$~\cite{combined}, as seen in Fig.~\ref{fig:nsr},
where we also see that $n_s \sim 0.97$. We also see in Fig.~\ref{fig:nsr} that monomial
single-field potentials $\propto \phi^n$ are disfavoured by the data, whereas the original model
of Starobinsky with an $R + R^2$ gravitational action~\cite{Staro} is highly compatible with the data,
as would be inflation driven by the Higgs field~\cite{Higgs} (with a non-mimimal coupling to gravity)
or in no-scale supergravity models~\cite{noscale}, as discussed later. We also note that the experimental
restriction on $n_s$, in particular, starts to provide an interesting constraint on the number
of e-folds of inflationary expansion, $N_*$, which is sensitive to how the inflaton decayed
into conventional matter~\cite{inflatondecay}, as also discussed later.

\subsection{Challenges for Inflationary Models}

It is already a challenge to write down a simple model of inflation that is compatible with the CMB data,
with simple power-law potentials being excluded as seen in Fig.~\ref{fig:Planck2015}, but there are other, deeper challenges.

How can one link inflation to `low-energy' physics at colliders? Within the Standard Model
of particle physics, the only candidate for the inflaton is the Higgs boson~\cite{Higgs}, but naive extrapolation
of Standard Model measurements to high energies indicates that the Higgs potential probably
turns negative~\cite{negative}, which is unsuitable for inflation.

Can one link inflation to some other well-motivated extension of the Standard Model?
For example, could the inflaton be a supersymmetric partner of a right-handed (singlet)
neutrino field~\cite{sneutrino}, or some sort of axion~\cite{natural}?

Can one link inflation to Planck-scale physics? Specifically, are there plausible inflatons
in string theory, or in compactified string models?

\subsection{The Starobinsky Model}

Let us first review the Starobinsky model~\cite{Staro}, which is based on a non-minimal action for general relativity:
\begin{equation}
S \; = \; \frac{1}{2} \int d^4 x \sqrt{-g} \left( R + \frac{R^2}{6 M^2} \right) \, ,
\label{RR2}
\end{equation}
where $M$ is some mass parameter that is unknown {\it a priori}.
It would seem superficially that the action (\ref{RR2}) does not contain a scalar
inflaton field. However, when one makes a conformal transformation to the Einstein frame~\cite{StelleWitt}
(in which the dependence on $R$ is purely linear) one find a scalar field model with a
characteristic potential:
\begin{equation}
S \; = \; \frac{1}{2} \int d^4 x \sqrt{{\tilde g}} \left[ {\tilde R} + (\partial_\mu \phi)^2 - \frac{3}{2} M^2 \left(1 - e^{-\sqrt{\frac{2}{3}} \phi} \right)^2 \right] \, .
\label{StelleWitt}
\end{equation}
Scalar CMB perturbations were first calculated in this model by~\cite{MC}, and
their small magnitude requires $M \sim 10^{13}~{\rm GeV} \ll M_P$, i.e., a
coefficient of the $R^2$ term in (\ref{RR2}) that is $\gg 1$ in natural units. The value of the scalar tilt $n_s$
and the magnitude of the tensor-to-scalar ratio $r \sim 3 \times 10^{-3}$ are highly compatible with the available
CMB data, as already seen in Fig.~\ref{fig:nsr}.

\subsection{Higgs Inflation}

William of Occam might have suggested using the (only established) Higgs scalar $h$ as the inflaton,
which would require a non-minimal coupling to gravity~\cite{Higgs}:
\begin{equation}
S \; = \; \frac{1}{2} \int d^4x \sqrt{-g} \left[ - (1 + \xi h^2) R + (\partial_\mu h)^2 -\frac{\lambda}{2} ( h^2 - v^2)^2 \right] \, ,
\label{hinf}
\end{equation}
where $\xi$ is an unknown parameter that must be $\gg 1$, $\lambda$ is the quartic Higgs self-coupling
and $v$ is the Higgs vev. When $1 \ll \xi \ll 10^{17}$, upon transforming to the Einstein frame one finds
\begin{equation}
S \; = \; \frac{1}{2} \int d^4 x \sqrt{- \tilde g} \left[ {\tilde R} + (\partial_\mu \chi)^2 - \frac{\lambda}{2 \xi^2} \left( 1 + e^{- \sqrt{\frac{2}{3}} \chi}\right)^{-2} \right] \, .
\label{hinfE}
\end{equation}
The resulting effective potential gives successful inflation when $\chi \gg 1$, where the potential is similar to that of
the Starobinsky model (though clearly different at small $\chi$).

The required value of $\xi \sim 15000$, and one may ask how natural that is. However, a more serious practical problem is that
one needs $\lambda > 1$ for field values beyond the Planck scale, whereas the best available calculations indicate
that probably $\lambda < 0$ because of renormalization by the top quark~\cite{negative}. One accurate calculation indicates
that $\lambda$ turns negative at a scale $\Lambda$:
\begin{equation}
\log \frac{\Lambda}{\rm GeV} \; =\; 11.3 + 1.0 \left( \frac{M_h}{\rm GeV} - 125.66 \right) - 1.2 \left( \frac{m_t}{\rm GeV} - 173.10 \right)
+ 0.4 \left( \frac{\alpha_3 (M_Z) - 0.1184}{0.0007} \right) \, .
\label{Lambda}
\end{equation}
Inserting the current experimental values of the Higgs mass $M_h = 125.09 \pm 0.24$~GeV~\cite{mh}, the top mass $m_t = 173.34 \pm 0.27 \pm 0.71$~GeV~\cite{mt}
and the strong-interaction coupling $\alpha_3 (M_Z) = 0.1177 \pm 0.0013$~\cite{Siggi}, one estimates that the Higgs self-coupling
$\lambda$ turns negative at $\ln (\Lambda/{\rm GeV}) = 10.0 \pm 1.0$. 

Negative $\lambda$ leads to metastability of the electroweak vacuum and the problem that most of the early universe would
have been sucked into the region where the potential is negative~\cite{Crunch}. Physics beyond the Standard Model is needed to
avoid these issues, and supersymmetry is one of the possibilities.

\subsection{Inflation Cries out for Supersymmetry and (No-Scale) Supergravity}

There are many other motivations for supersymmetry, and inflation is one of them~\cite{cries}.
Models of inflation involve a scalar inflaton field that is elementary, at least at energies $\ll M_P$,
and obtaining the right magnitude of perturbations requires a mass $\ll MP$ (typically $\sim 10^{13}$~GeV)
and/or a small inflaton self-coupling $\lambda \ll 1$. Both of these are technically natural in
supersymmetric theories.

The first supersymmetric models of inflation were based on global supersymmetry. However,
experience (general relativity, the Standard Model) teaches us that the only good symmetry is a local symmetry,
and the same should apply to supersymmetry. The local version of supersymmetry necessarily involves
local space-time coordinate transformations, and is supergravity. Clearly also, any discussion of
early Universe cosmology needs gravity, and supergravity is the way to combine gravity and supersymmetry.

However, there is an issue with generic supergravity models when matter is included, which is that
their effective potentials typically have `holes' with depths $\sim {\cal O}(M_P^2)$. These make any
treatment of cosmology problematic, let alone inflation. There is, however, an exception, namely
the class of no-scale supergravity models~\cite{oldnoscale}. The effective potentials in this class of supergravity models
have flat directions, and the scalar potential resembles that in globally supersymmetric models, with
controlled corrections. 

An added motivation for considering no-scale supergravity models is that they arise naturally
in compactifications of string theory~\cite{Witten}. Therefore, no-scale models of inflation may offer some
prospects for making contact with some underlying variant of string theory.

\subsection{No-Scale Supergravity Models of Inflation}

The first no-scale supergravity models of inflation appeared in the 1980s~\cite{GL, gang}, largely motivated by the
`unholy' form of the effective potential. This approach was revived in 2013~\cite{noscale} in response to the new,
high-precision CMB data from Planck. Supergravity models are characterised by a Hermitian
K\"ahler potential $K$ that specifies the internal geometry of the scalar fields $\phi$, whose kinetic
terms are $(\partial^2 K /\partial \phi^i \partial \phi^*_j) \partial_\mu \phi^i \partial^\mu \phi^*_j$. The simplest
no-scale inflationary model contains two complex scalar fields:
\begin{equation}
K \; = \; - 3 \ln \left( T + T^* - \frac{|\phi|^2}{3} \right) \, ,
\label{noscaleK}
\end{equation}
and the potential interactions between the scalar fields are characterised by a superpotential $W$,
which we assumed to have the form
\begin{equation}
W \; = \; \frac{\mu}{2} \phi^2 - \frac{\lambda}{3} \phi^3 \, ,
\label{WZW}
\end{equation}
as first studied by Wess and Zumino~\cite{WZ}. We assumed that the field $T$, which could be interpreted
as a modulus of string compactification, is fixed by string dynamics: $T = c/2$, in which case the
effective Lagrangian for the inflaton field $\phi$ is
\begin{equation}
{\cal L}_{eff} \; = \; \frac{c}{(c - |\phi|^2/3)^2} |\partial_\mu \phi|^2 - \frac{\hat V}{(c - |\phi|^2/3)^2}: \; 
{\hat V} \; = \; \left| \frac{\partial W}{\partial \phi} \right|^2 \, .
\label{Leff}
\end{equation}
We note that the prefactors in the Lagrangian terms in ({\ref{Leff}) modify the form of the
effective potential that one would have obtained in the globally-supersymmetric case,
when one transforms to a canonically-normalized inflaton field.

The Wess-Zumino superpotential gives good inflation already in the global case~\cite{CEM}, and we
found this to be possible also in its no-scale supergravity incarnation for $\lambda/\mu \sim 1/3$~\cite{noscale},
as seen in Fig.~\ref{fig:noscaleStaro}. Indeed, for the specific case $\lambda = \mu/3$, shown
as the black line in Fig.~\ref{fig:noscaleStaro}, the effective scalar potential is {\it identical}
with that in the original Starobinsky model (\ref{RR2}, \ref{StelleWitt}).

\begin{figure}[h!]
\centering
\vskip -3cm
		\scalebox{0.4}{\includegraphics{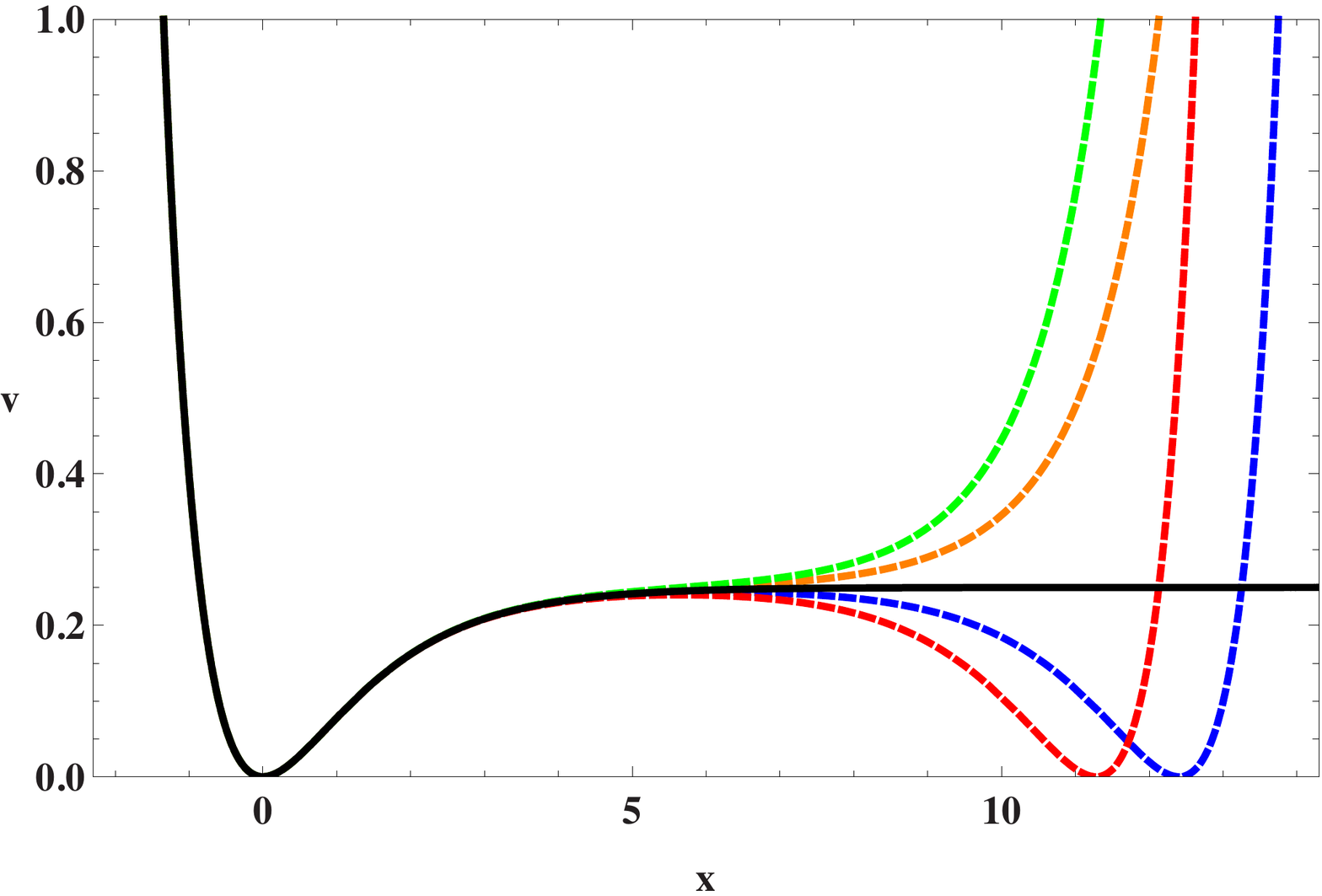}} 
\vskip -3cm
\caption{The inflationary potential $V$ in the Starobinsky $R + R^2$ model~\protect\cite{Staro} (solid black line) compared with its form in
various no-scale models~\protect\cite{noscale} (dashed coloured lines).}
\label{fig:noscaleStaro}
\end{figure}

Looking beyond the Starobinsky limit, one can consider generic potentials of the form~\cite{ENO7}
\begin{equation}
V \; = \; A \left( 1 - \delta e^{-B\phi } + {\cal O}(e^{-2 B \phi}) \right) \, .
\label{beyondS}
\end{equation}
In such models, the most prominent inflationary predictions are
\begin{equation}
n_s \; = \; 1 - 2 B^2 \delta e^{B \phi_*}, \; r \; = \; 8 B^2 \delta^2 e^{-2 B \phi_*} , \; N_* = \frac{1}{B^2 \delta}e^{B \phi_*} \, ,
\label{generalS}
\end{equation}
where $\Phi_*$ is the value of the inflaton field that yields $N_*$ e-folds of inflation.
It is clear from (\ref{generalS} that $n_s$ is in a one-to-one relationship with $N_*$,
and that $r$ may be adjusted, for any fixed value of $N_*$, by varying the coefficient $B$
in the exponential in (\ref{beyondS})~\footnote{Similar results are found in the so-called $\alpha$-attractor models~\cite{KL}.}.

As an example how this freedom could be exploited, one can consider models in which
the no-scale property is shared among several moduli fields~\cite{ENO7}:
\begin{equation}
K \; \ni \; - \sum_i \ln ( T_i + T_i^*): \; N_i > 0, \; \sum_i N_i \; = \; 3 \, ,
\label{manyT}
\end{equation}
as is characteristic of generic string compactifications. In this case,
if the inflaton is identified with $T_i$ one has
\begin{equation}
B \; = \; \sqrt{\left(\frac{2}{N_i}\right)}, \; r \; = \; \frac{4 N_i}{N_*^2} \, ,
\label{generalB}
\end{equation}
with the specific choice of a single modulus with $N_i =3$ corresponding to
the Starobinsky model ({\ref{RR2}, \ref{StelleWitt}). A measurement of $r$ could
therefore, within this no-scale supergravity framework, provide a unique window
on the phenomenology of string compactification.

\subsection{How Many e-folds of Inflation?}

The discussion above shows that the number of e-folds, $N_*$, is a key to the
phenomenology of inflationary models, and we also see in Fig.~\ref{fig:nsr} that
the CMB data are beginning to provide interesting constraints on $N_*$~\cite{inflatondecay}. This
can be related to the potential $V_*$ at $\phi_*$ corresponding to the scale $k_*$:
\begin{equation}
N_* \; = \; 67 - \ln\left(\frac{k_*}{a_0 H_0} \right) + \frac{1}{4} \ln \left( \frac{V_*^2}{M_P^2 \rho_{end}} \right)
+ \frac{1 - 3 w_{int}}{12(1 + w_{int})} \ln \left( \frac{\rho_{reh}}{\rho_{end}} \right) - \frac{1}{12} \ln g_{th} \, ,
\label{N*}
\end{equation}
where $\rho_{end, reh}$ are the densities at the end of inflation and after reheating,
respectively, $w_{int}$ is the average equation-of-state parameter during inflaton
decay and reheating, and $g_{th}$ is the number of degrees of freedom after reheating.

\subsection{Constraints on Inflaton Decay}

The value of $N_*$ and hence $r$, in particular, is sensitive to the inflaton decay rate
$\Gamma_\phi$, since the ratio $\rho_{reh}/\rho_{end} \propto \Gamma_\phi^2$.
Thus a measurement of $r$ can constrain models of inflaton decay. This point is
illustrated in Fig.~\ref{fig:N*WZ} for the Starobinsky-like no-scale supergravity models
of inflation whose potentials are shown in Fig.~\ref{fig:noscaleStaro}. In particular, for
the special case $\lambda/\mu = 1/3$ that reproduces exactly the Starobinsky model,
we see that $N_* \ge 44$ at the 95\% CL and $\ge 50$ at the 68\% CL. If the dominant
inflaton decay is via a trilinear superpotential coupling $y$ into two-body final states, the inflaton
decay rate $\Gamma_\phi = |y|^2/8 \pi m_\phi$, and these constraints on $N_*$ correspond
to the constraints $y > 10^{-16}$ at the 95\% CL and $> 10^{-7}$ at the 68\% CL~\cite{inflatondecay}. 

\begin{figure}[h!]
\centering
\vspace{-4cm}
\scalebox{0.45}{\includegraphics{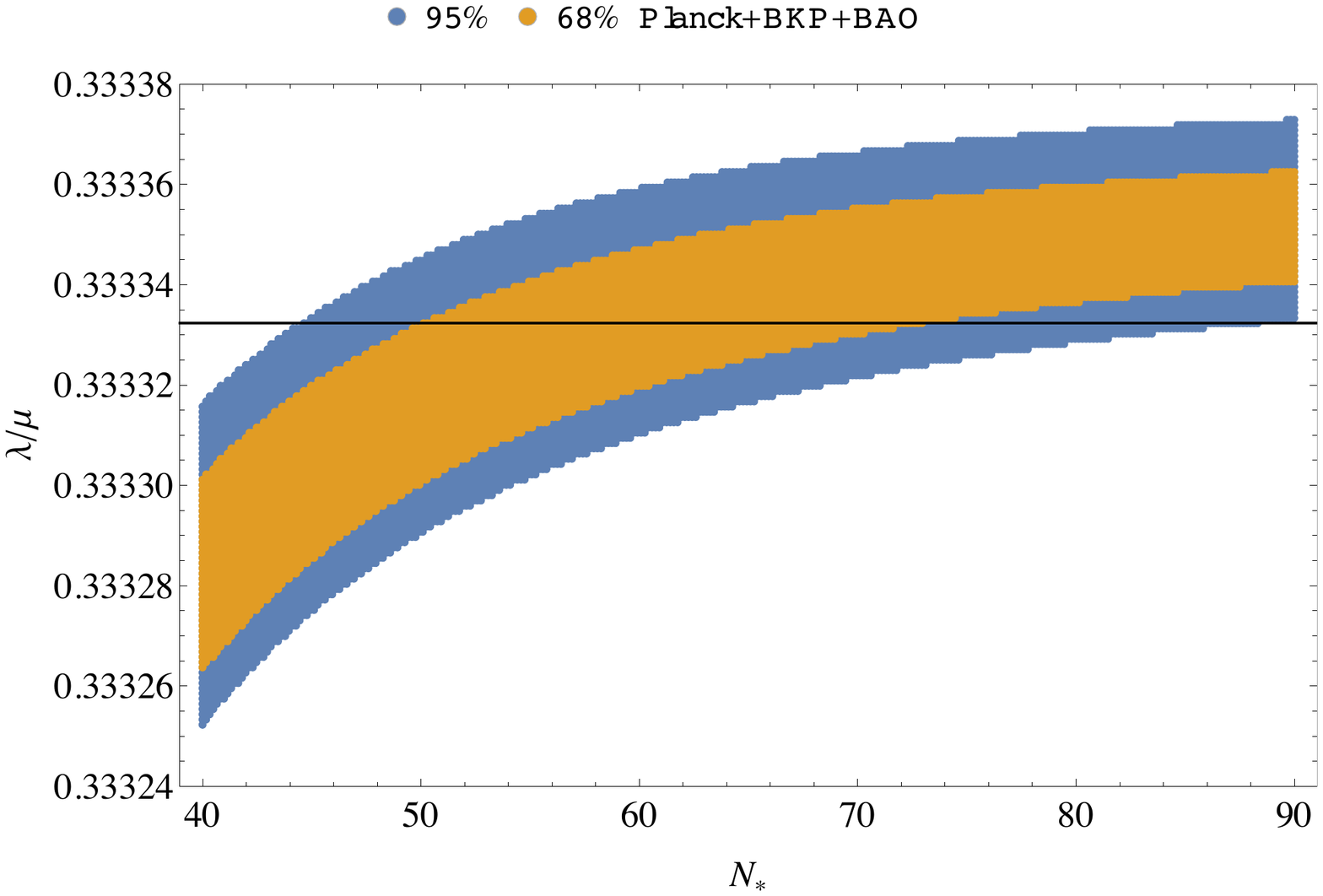}}
\vspace{-3.5cm}
	\caption{The 68\% and 95\% CL regions (yellow and blue, respectively) in the $(N_*, \lambda/\mu)$ plane
	for the no-scale inflationary model~\protect\cite{noscale} with a matter inflaton field and the Wess-Zumino 
	superpotential~\protect\cite{inflatondecay}. The horizontal black line is for $\lambda/\mu = 1/3$, the value that reproduces the
	inflationary predictions of the Starobinsky model~\protect\cite{noscale}.} 
	\label{fig:N*WZ}
\end{figure} 

These bounds are to be compared with those due to upper limits on $y$ from the
production in inflaton decays of gravitinos with mass $m_{3/2}$~\cite{inflatondecay}. We find
\begin{equation}
\vert y \vert \; \lesssim \;  1.5 \times 10^{-6} \quad {\rm for} \quad m_{3/2} = 6~{\rm TeV} \, ,
\label{ybounds}
\end{equation}
from requiring that long-lived gravitinos do not mess up the success of standard
Big-Bang nucleosynthesis (BBN) calculations, weakening for larger $m_{3/2}$ and disappearing altogether
if the gravitino is sufficiently heavy to decay before BBN. There is also a constraint
\begin{equation}
\vert y \vert \; < \; 2.7 \times 10^{-5} \times \left( \frac{100 {\rm GeV}}{m_{\rm LSP}} \right) \, ,
\label{ymLSP}
\end{equation}
from the upper limit on the contribution of the lightest supersymmetric particle
(LSP) to the cosmological cold dark matter. The latter constraint, in particular,
suggests that $\vert y \vert \lesssim 10^{-5}$ for LSP masses in the range of a few hundred GeV,

The constraint on the inflaton decay coupling $y$ from the CMB measurement of $r$
is not yet competitive with the other constraints (\ref{ybounds}, \ref{ymLSP}), but this
can be expected to change in the future.

\section{Meanwhile, back at the LHC}

During 2015, the LHC restarted operations, making collisions at a centre-of-mass
energy of 13~TeV, much higher than the 7 and 8~TeV during its first run. The two
experiments looking for new high-mass physics, ATLAS and CMS, have each accumulated
$\sim 3$/fb of high-energy data. They presented their first preliminary data during this
conference, and created a sensation.

Both experiments reported statistical excesses in their $\gamma \gamma$ invariant-mass
distributions that could perhaps, maybe, be hints of a possible new particle $X$
with mass $\sim 750$~GeV decaying into two photons~\cite{X}. The significances in the experiences
are only $\sim 3$~$\sigma$, and are diluted by the `look-elsewhere' effect that the
experiments examined many $\gamma \gamma$ invariant-mass distributions
(and many other distributions!), so the overall significance falls far short of the 5-$\sigma$
`gold standard' required to claim discovery of a new particle, particularly one so
unexpected. If it exists, the $X$ particle would herald a layer of new physics beyond the
Standard Model.

\subsection{Interpretation of the $X(750)$ Hint}

The CMS and ATLAS data suggest a product of cross-section and $\gamma \gamma$
branching ratio $\sim 6$/fb, and most analyses assume that it is a scalar or pseudoscalar.
The most plausible possibility is that it decays to $\gamma \gamma$ via triangular loops 
of heavy charged particles, most likely fermions weighing $\gtrsim 400$~GeV~\cite{EEQSY}. The
production of the $X(750)$ is also often thought to be via loops of massive coloured fermions. These
could not be a fourth generation of Standard Model particles, and minimal supersymmetric
models are also unable to explain the magnitude of the $X$ `signal'. There would have to be a
whole new world out there!

Four simple possibilities for these extra fermions are: i) a single vector-like charge 2/3 quark,
ii) an isospin doublet of vector-like charge 2/3 and -1/3 quarks, iii) an isodoublet and two
singlet charge 2/3 and -1/3 quarks, and iv) a complete vector-like generation including leptons
as well as quarks~\cite{EEQSY}. Fig.~\ref{fig:Xcoupling} illustrates the possible couplings $\lambda$
in these four models of $X$
to the vector-like fermions (assumed, for simplicity, to be universal) as functions of their
masses (also assumed, for simplicity, to be universal) if they are to reproduce the possible
`signal' (assuming that the dominant $X$ decays are those mediated by fermion loops).
It is worth noting that model iv) contains a weakly-interacting neutral fermion that is a
possible dark matter candidate.

\begin{figure}[!h]
 \centerline{\includegraphics[scale=0.35]{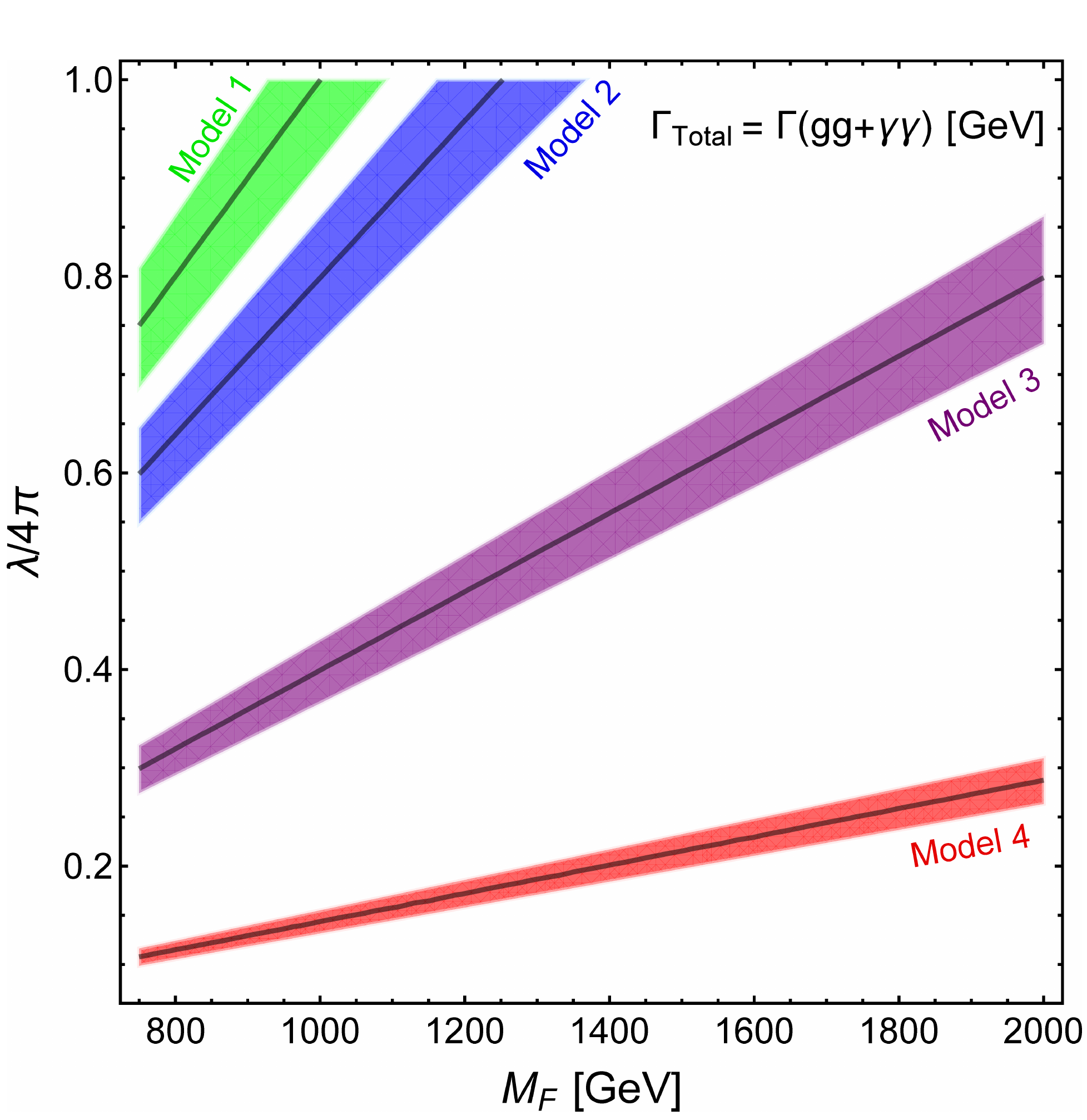}}
\vspace*{0.1cm}
\caption{Values of the vector-like fermion mass $m_F$ and coupling $\lambda$ (both assumed to be universal) required models i), ii), iii) and iv) to accommodate the possible $gg\to \Phi \to \gamma \gamma$ signal reported by CMS and ATLAS~\protect\cite{X}. The black lines are for the central value of the cross section, $6$~fb, and the coloured bands represent the $1$-$\sigma$ uncertainties. Figure
from~\protect\cite{DEGQ}, adapted from~\protect\cite{EEQSY}.}
\label{fig:Xcoupling}
\vspace*{-.2mm}
\end{figure}

In each of these models, triangular diagrams generate decays of the $X$ particle to
other pairs of vector bosons: gluon-gluon, $Z \gamma$, $ZZ$ and $W^+ W^-$, at rates relative to
$\gamma \gamma$ that are easily calculable and depend on the model~\cite{EEQSY}.
It is premature to exclude any of these models, though model ii) is under pressure from searches
at the LHC. In addition to these decays, the LHC experiments should also look for the heavy
fermions that are required. There will be theoretical and experimental work for a generation,
{\it if} the $X(750)$ exists. Since the LHC should make many more collisions in 2016, 
we can expect to know soon whether the $X(750)$ `signal' is real.
In the mean while, we theorists are ``allowed to be a little excited"~\cite{FG}~\footnote{However, we should also
remember the words of Laplace, ``Plus un fait est extraordinaire, plus il a besoin d'{\^e}tre appuy\'e de fortes preuves."
(``The more extraordinary a claim, the stronger the proof required to support it.")}.

\subsection{What's in it for Inflation?}

The appearance of another scalar particle would provide another candidate for the inflaton,
and one can consider inflation via a non-minimal $X$ coupling to gravity~\cite{Xinflation}, just as was discussed
for Higgs inflation in Subsection 2.4. The appearance of the $X$ particle could also be used to
revive Higgs inflation, or even lead one to consider a two-field model of inflation. If it is real, the $X(750)$
particle could revolutionise inflationary model-building as well as collider physics.

\section{Summary}

As we saw in Section 2, many popular models of inflation postulate some modification of the
minimal Einstein-Hilbert action: maybe an $R^2$ term as in the Starobinsky, maybe a
non-minimal coupling to a scalar field as in Higgs (or $X$!) inflation, or maybe promote
general relativity to supergravity. It is ironic that, 100 years after the proposal of general
relativity, at a time when experiment finally confirms its key prediction of gravitational waves,
the time may have come to move on from Einstein. As we saw in Section~3, if the possible
hint, maybe, of a new particle $X$ with mass $\sim 750$~GeV would, if confirmed,
require us to move on from the Standard Model of particle physics, potentially also with
interesting implications for inflationary model-building.

\section*{Note added}

The discovery of gravitational waves~\cite{GW}, which was announced while this write-up was being completed,
is a tremendous vindication of Einstein's theory of general relativity~\cite{GWgr}. Perhaps this and similar future observations
will give us hints how to extend it? In the mean time, the gravitational-wave data already provide some constraints on
possible modifications of Einstein's theory, including an upper limit on the graviton mass~\cite{GWgr} $\sim 10^{-22}$~g, 
limits on Lorentz violation~\cite{EMN, KM} and the violation of the principle of equivalence~\cite{PE}, 
and the possible observation of a $\gamma$-ray flash (if confirmed) would
tells us that electromagnetic and gravitational waves travel at the same velocity~\cite{EMN, 17} to within $10^{-17}$.
{\it Une affaire \`a suivre!}

\section*{Acknowledgments}

The author was supported partly by the London Centre for Terauniverse Studies (LCTS), 
using funding from the European Research Council via the Advanced Investigator Grant 26732, 
and partly by the STFC Grant ST/L000326/1. The author thanks Pisin Chen and his team for
their kind hospitality during the Second LeCosPA Symposium, and thanks the Universidad de Antioquia, Medell\'in, for its
hospitality while writing up this talk, using grant FP44842-035-2015 from Colciencias (Colombia).


\end{document}